\documentclass[useAMS,usenatbib,usegraphicx,usedcolumn]{mn2e}

\usepackage{times} %***
\usepackage{balance} %***

\defcitealias{2007MNRAS.382..109T}{T07}

\newcommand{\Mdyn}{\ensuremath{M_\mathrm{dyn}}}
\newcommand{\Mstar}{\ensuremath{M_\star}}
\newcommand{\Msun}{\ensuremath{\mathrm{M_\odot}}}

\newcommand{\re}{\ensuremath{r_\mathrm{e}}}
\newcommand{\rShen}{\ensuremath{r_\mathrm{Shen}(M_\star)}}
\newcommand{\kpc}{\ensuremath{\mathrm{kpc}}}

\newcommand{\sigmae}{\ensuremath{\sigma_\mathrm{e}}}
\newcommand{\kms}{\ensuremath{\mathrm{km \  s^{-1}}}}
\title[Dynamical-stellar mass discrepancy]{The discrepancy between dynamical
and stellar masses in massive compact galaxies traces non-homology}

\author[L. Peralta de Arriba et al.]{Luis
Peralta de Arriba,$^{1,2}$\thanks{E-mail: lperalta@iac.es}
Marc Balcells,$^{1,2,3}$
Jes\'us Falc\'on-Barroso$^{1,2}$
\newauthor and Ignacio Trujillo$^{1,2}$\\
$^{1}$Instituto de Astrof\'{\i}sica de Canarias (IAC),
      E-38200 La Laguna, Tenerife, Spain\\
$^{2}$Universidad de La Laguna, Deparment Astrof\'{\i}sica,
      E-38206 La Laguna, Tenerife, Spain\\
$^{3}$Isaac Newton Group of Telescopes,
      E-38700 Santa Cruz de La Palma, La Palma, Spain}

\begin{document}

\date{Accepted 2014 February 14. Received 2014 February 14;
in original form 2013 July 17}
\setcounter{page}{1634} %***
\volume{440} %***

\pagerange{\pageref{firstpage}--\pageref{lastpage}} \pubyear{2014}

\maketitle

\label{firstpage}

\begin{abstract}
For many massive compact galaxies, their dynamical masses ($\Mdyn \propto \sigma^2 \re$) are lower than their stellar masses (\Mstar). We analyse the unphysical mass discrepancy $\Mstar / \Mdyn > 1$ on a stellar-mass-selected sample of early-type galaxies ($\Mstar \ga 10^{11} \  \Msun$) at redshifts $z$ $\sim$ 0.2 to $z$ $\sim$ 1.1. We build stacked spectra for bins of redshift, size and stellar mass, obtain velocity dispersions, and infer dynamical masses using the virial relation $\Mdyn \equiv K \  \sigmae^2 \re / G$ with $K$ = 5.0; this assumes homology between our galaxies and nearby massive ellipticals. Our sample is completed using literature data, including individual objects up to $z$ $\sim$ 2.5 and a large local reference sample from the Sloan Digital Sky Survey (SDSS). We find that, at all redshifts, the discrepancy between \Mstar\ and \Mdyn\ grows as galaxies depart from the present-day relation between stellar mass and size: the more compact a galaxy, the larger its $\Mstar / \Mdyn$. Current uncertainties in stellar masses cannot account for values of $\Mstar / \Mdyn$ above 1. Our results suggest that the homology hypothesis contained in the \Mdyn\ formula above breaks down for compact galaxies. We provide an approximation to the virial coefficient $K \sim 6.0 \left[ \re / (3.185 \  \kpc) \right]^{-0.81} \left[ \Mstar / (10^{11} \  \Msun) \right]^{0.45}$, which solves the mass discrepancy problem. A rough approximation to the dynamical mass is given by $\Mdyn \sim \left[ \sigmae / (200 \  \kms) \right]^{3.6} \left[ \re / (3 \  \kpc) \right]^{0.35} 2.1 \times 10^{11} \  \Msun$.
\end{abstract}


\section{Introduction}

Observations in the last decade have shown that the mean size of massive ($\Mstar \ga 10^{11} \  \Msun$) galaxies evolves with redshift ($z$); that is, these galaxies have, at a fixed stellar mass, a smaller size at higher redshifts \citep[e.g.][]{2005ApJ...626..680D,2006MNRAS.373L..36T,2007MNRAS.374..614L,2007ApJ...671..285T,2007ApJ...656...66Z,2008ApJ...687L..61B,2008A&A...482...21C}. The size evolution is more dramatic for early-type galaxies (ETGs) than for late-type galaxies \citep[][hereinafter \citetalias{2007MNRAS.382..109T}]{2007MNRAS.382..109T}. In fact, at $z$ $\sim$ 2 a large number of massive ETG galaxies have sizes, as parametrized by the effective radius, of $\sim$1 kpc. These objects have been termed massive compact galaxies in the literature. Only a few of these galaxies have been found in the nearby Universe \citep[e.g.][]{2009ApJ...692L.118T,2010ApJ...720..723T,2013ApJ...762...77P,2014ApJ...780L..20T}.

With the increasing availability of velocity dispersion measurements for massive compact galaxies, it has become apparent that dynamical masses, estimated using the virial relationship $\Mdyn \propto \sigma^2 \re$, often turn out to be lower than stellar masses. This occurs at low as well as at high redshifts. \citet{2010ApJ...709L..58S} and \citet{2011ApJ...738L..22M} found this problem for two galaxies at $z$ $\sim$ 0.5 and for four galaxies at $z$ $\sim$ 1, respectively. \citet{2012MNRAS.423..632F} report the same result for seven massive compact galaxies in the nearby Universe ($z$ $\sim$ 0.14).

As the inequality $\Mstar > \Mdyn$ is unphysical, it follows that, for these objects, either the dynamical masses are underestimated or the stellar masses are overestimated. Both conditions could of course apply simultaneously.

Stellar mass determinations are based on the comparison of galaxy spectral energy distributions (SEDs) with synthetic spectra of stellar populations built using the current knowledge of stellar spectra, stellar evolution, the initial mass function (IMF), star formation history and dust attenuation. It is believed that the uncertainty in these parameters can lead to an error in the estimation of the stellar mass by factors of up to 2--4 \citep{2009ApJ...699..486C,2009ApJ...701.1839M}. The comparison with stellar synthesis models can be carried out using broad-band photometric data, and hence this method provides an efficient means of estimating stellar masses for galaxies at high-redshift.

Several techniques are available to derive dynamical masses. The most accurate methods are based on one of two approaches: the solution of Poisson and Jeans equations \citep[e.g.][]{1980PASJ...32...41S}, or the description of the system using an orbit-superposition method \citep{1979ApJ...232..236S}. These methods require high-quality two-dimensional spectroscopic data, making them expensive in observing time. A cheaper alternative, and the only one available today for high-redshift galaxies, is to use simpler mass estimators based on the virial theorem. Following the notation in \citet{1988ASPC....4..329D}, we write the virial theorem as
\begin{equation} \label{eq:virial}
\frac{G \Mdyn}{\langle r \rangle} = k_E \  \frac{\langle v^2 \rangle}{2},
\end{equation}
where $G$ is the gravitational constant, $\langle r \rangle$ the gravitational radius of the system, $\langle v^2 \rangle$ is twice the kinetic energy of the system per unit mass, and $k_E$ is the virialization constant. $k_E$ has the value $k_E = 2$ when the system is virialized. Unfortunately, $\langle r \rangle$ and $\langle v^2 \rangle$ are not direct observables. For this reason, it is common to define two coefficients, $k_r$ and $k_v$, which connect these magnitudes with direct observables that provide galactic length and velocity scales respectively ($O_r$ and $O_v$). These coefficients are defined by the following equalities:
\begin{equation} \label{eq:k_r}
\langle r \rangle = k_r O_r,
\end{equation}
\begin{equation} \label{eq:k_v}
\langle v^2 \rangle = k_v O_v^2.
\end{equation}
$k_r$ and $k_v$ are coefficients describing the spatial and dynamical structure of the objects, respectively. Using equations~(\ref{eq:k_r}) and (\ref{eq:k_v}), it is straightforward to express the virial theorem (equation~\ref{eq:virial}) in terms of observables:
\begin{equation} \label{eq:virial-obs-gen}
\Mdyn = K \  \frac{O_v^2 O_r}{G},
\end{equation}
where we have defined $K$ as
\begin{equation} \label{eq:k}
K \equiv \frac{k_E}{2} k_v k_r.
\end{equation}
The coefficient $K$ will be a universal constant only if the coefficients $k_r$ and $k_v$ are the same for all ETGs (or if their dependences disappear in their product). This hypothesis is known as homology, because it would be verified if all ETGs had the same mass density, and kinematic and luminosity structures. Following these ideas, \citet{2006MNRAS.366.1126C} selected the effective (half-light) radius \re\ and the luminosity-weighted second moment \sigmae\ of the line-of-sight velocity distribution (LOSVD) within \re\ to play the role of observables $O_r$ and $O_v$; that is, they used the following equation:
\begin{equation} \label{eq:virial-rsigma}
\Mdyn = K \  \frac{\sigmae^2 \re}{G}.
\end{equation}
Using a sample of nearby and normal-sized ETGs, these authors found that $K$ is approximately constant and they proposed the following calibration as a reliable estimator of the dynamical mass for an ETG:
\begin{equation} \label{eq:virial-cap}
\Mdyn = (5.0 \pm 0.1) \  \frac{\sigmae^2 \re}{G}.
\end{equation}
This formula is used in the references mentioned above that lead to the reported discrepancy between dynamical and stellar masses. Hereinafter, we will always use the term dynamical mass to refer to the mass calculated using equation~(\ref{eq:virial-cap}).

In this work we characterized the fraction $\Mstar / \Mdyn$ at various redshifts and covering various regions of the mass--size space. We used this information to determine in which areas of this parameter space the mass predictions produce the impossible result $\Mstar > \Mdyn$.

Our paper is structured as follows. In Section~\ref{sec:data}, we present a description of our sample and additional sources of data. We describe the processing of spectra and velocity dispersion measurements of our sample in Section~\ref{sec:processing}. In Section~\ref{sec:discrepancy}, we show our results for $\Mstar / \Mdyn$ and argue that \Mstar\ uncertainties alone cannot explain values of $\Mstar / \Mdyn > 1$. We interpret the discrepancy as a non-homology effect in Section~\ref{sec:interpretation}. In Section~\ref{sec:discussion} we discuss about our results. We summarize our conclusions in Section~\ref{sec:conclusions}. We adopt a $\Lambda$CDM cosmology with $\Omega_\mathrm{m} = 0.3$, $\Omega_\Lambda = 0.7$ and $H_0 = 70 \ \mathrm{km \ s^{-1} \ Mpc^{-1}}$. The stellar masses are obtained assuming a Salpeter IMF (we applied a conversion factor when necessary following the prescriptions from \citealt{2009MNRAS.394..774L}).

\section{Description of the data} \label{sec:data}

\subsection{Cross-matching T07 massive galaxies with the DEEP2 DR4 survey} \label{subsec:t07deep2-cross-matching}

Our targets belong to the catalogue of massive galaxies studied by \citetalias{2007MNRAS.382..109T}. This catalogue contains 796 galaxies with redshifts between 0.2 and 2. The authors split their sample into two groups: one containing objects that have a S\'ersic index $n$ $<$ 2.5 (disc-like galaxies) and the other containing galaxies that have $n$ $>$ 2.5 (spheroid-like galaxies). This criterion is based on the correlation between the S\'ersic index $n$ and the morphological type observed both in the nearby Universe \citep{2004ApJ...604L...9R} and in the high-redshift Universe \citep[see e.g.][]{2013MNRAS.428.1460B}. In this work, we focus on the spheroid-like galaxies, which are the set with a stronger size evolution. Therefore, our initial catalogue contains the 463 massive galaxies with $n$ $>$ 2.5.

An extra requirement we apply to this subsample is the availability of a spectrum in the unique redshift DEEP2 DR4 catalogue \citep{2003SPIE.4834..161D,2007ApJ...660L...1D,2013ApJS..208....5N}. DEEP2 DR4 was designed to conduct a comprehensive census of massive galaxies, and their properties, environments and large-scale structure down to an absolute magnitude $M_B = -20$ at $z$ $\sim$ 1. The targets of this catalogue are located in four fields, which have a total area of 2.8 deg$^2$. One of these fields is the Extended Groth Strip \citep[EGS,][]{2007ApJ...660L...1D}. All the galaxies studied by \citetalias{2007MNRAS.382..109T} are located in the EGS field. This is relevant because the EGS is the only field of the DEEP2 DR4 survey in which colour pre-selection was not applied.

The cross-match between the DEEP2 DR4 catalogue and the spheroid-like \citetalias{2007MNRAS.382..109T} galaxies was performed using the \emph{OBJNO} index on the first and the \emph{galaxy identification} on the second. Both indices come from the sample defined by \citet{2004ApJ...617..765C}. The number of matches is 260.

Other requirements were applied to this set in order to avoid contaminant objects. Two targets were rejected for being too small to be reliable (they had a semimajor axis equal to 0.01 arcsec and an effective radius $\sim$0.01 kpc). We also required our galaxies to have a secure redshift, namely galaxies with quality codes equal to 3 or 4 in the DEEP2 DR4 catalogue. Only in two cases was the difference between the redshift used by \citetalias{2007MNRAS.382..109T} ($z_\mathrm{T07}$) and the secure redshift published by DEEP2 DR4 ($z$) was significant; that is, it was not the case that $|z_\mathrm{T07} - z| / (1 + z)$ $<$ 8 per cent, 0.89 $<$ $z / z_\mathrm{T07}$ $<$ 1.15 and $|z_\mathrm{T07} - z| / z$ $<$ 13 per cent. These two targets were rejected. With these extra requirements, our final catalogue comprised 243 sources.

\begin{figure*}
  \includegraphics{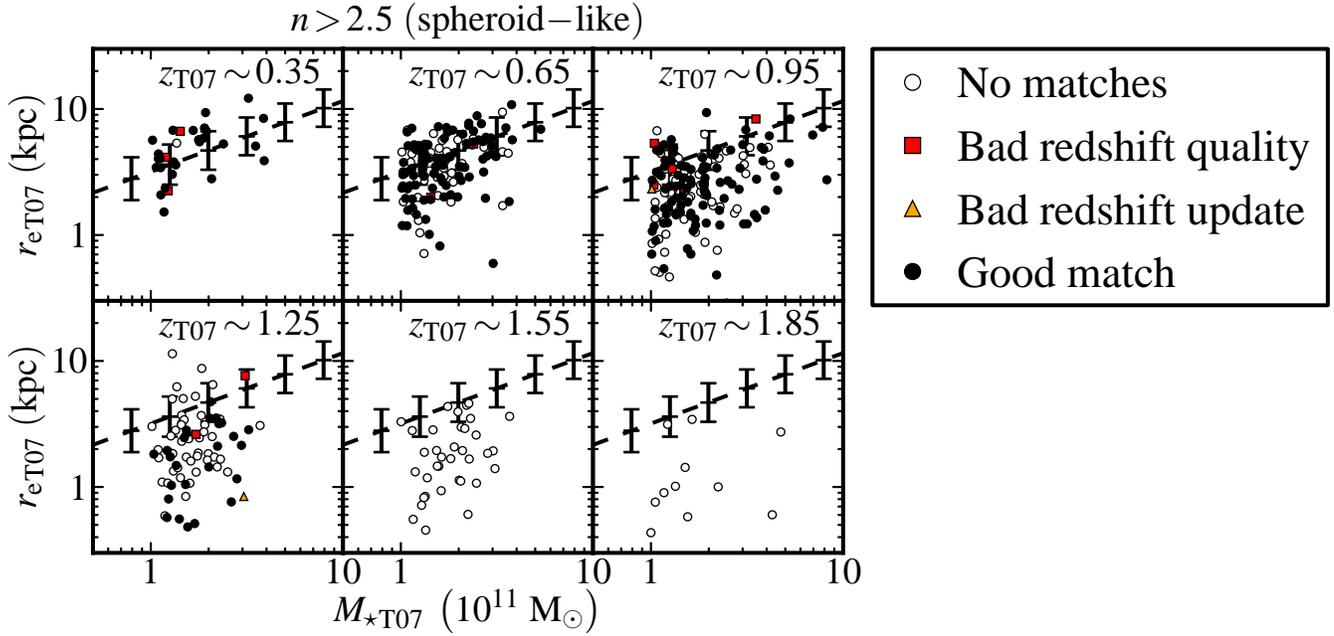}
  \caption{The stellar mass--size distribution of the spheroid-like galaxies ($n$ $>$ 2.5) of the \citetalias{2007MNRAS.382..109T} sample. The squares and triangles denote objects that are  discarded because their DEEP2 DR4 redshifts are not secure, and the differences between the redshift used by \citetalias{2007MNRAS.382..109T} and the secure redshift published by DEEP2 DR4 are significant (see text for more details). The black circles represent the matches between \citetalias{2007MNRAS.382..109T} and DEEP2 DR4 that we have not rejected, while the white circles are \citetalias{2007MNRAS.382..109T} galaxies without available spectra in the DEEP2 DR4 sample. The dashed line shows the relationship that the early-type galaxies follow in the nearby Universe \citep{2003MNRAS.343..978S}. Error bars over-plotted on the dashed line show the dispersion of this relationship. Stellar masses, radii and redshifts used in this figure are the same as those used in \citetalias{2007MNRAS.382..109T}.}
  \label{fig:t07deep2-cross-matching}
\end{figure*}

In Fig.~\ref{fig:t07deep2-cross-matching} we plot the information from the cross-match between the two catalogues in the stellar mass--size diagram. It can be seen that our sample is restricted to $z_\mathrm{T07}$ $<$ 1.4. In the three lower redshift bins, most of the points satisfy our requirements, but this is not the case in the $z_\mathrm{T07}$ $\sim$ 1.25 bin. A relevant fact that we can check in Fig.~\ref{fig:t07deep2-cross-matching} is that there are no biases affecting the stellar mass--size distribution in the three lower redshift bins.

\subsection{Information taken from T07 and DEEP2 DR4 catalogues} \label{subsec:t07deep2-magnitudes}

We summarize here the parameters that we will use throughout this work, detailing in each case whether the parameter was extracted from the DEEP2 DR4 or the \citetalias{2007MNRAS.382..109T} catalogue, or whether it was derived from a combination of parameters from both catalogues.

\begin{enumerate}
\item S\'ersic index $n$: taken from the \citetalias{2007MNRAS.382..109T} catalogue.
\item Redshift $z$: taken from the DEEP2 DR4 catalogue, so all values are spectroscopic.
\item Circularized effective radius \re: effective radii in arcseconds were obtained from \citetalias{2007MNRAS.382..109T}, and these quantities were then converted to kiloparsecs using the DEEP2 DR4 redshifts. We compared our effective radii in kiloparsecs with the \citetalias{2007MNRAS.382..109T} radii ($r_\mathrm{eT07}$), and checked that they were very similar, namely that they satisfied $0.93 < \re / r_\mathrm{eT07} < 1.05$.
\item Stellar mass \Mstar: this magnitude was recalculated based on the \citetalias{2007MNRAS.382..109T} stellar masses ($M_{\star\mathrm{T07}}$) and considering the DEEP2 DR4 redshift ($z$) as a small correction of \citetalias{2007MNRAS.382..109T} redshift ($z_\mathrm{T07}$). Specifically, we considered that the luminosity of an object should be corrected using the equation $L / L_\mathrm{T07} = [d(z) / d(z_\mathrm{T07})]^2$, where $L$ is the updated luminosity at redshift $z$, $L_\mathrm{T07}$ is the luminosity at \citetalias{2007MNRAS.382..109T} redshift $z_\mathrm{T07}$, and $d(z)$ is the luminosity distance at redshift $z$. We also assumed that the stellar mass--luminosity ratio should be very similar to that considered by \citetalias{2007MNRAS.382..109T} for each galaxy; that is, we used that $\Mstar / L = k = M_{\star\mathrm{T07}} / L_\mathrm{T07}$. These hypotheses imply that the mass correction formula is $\Mstar = [d(z) / d(z_\mathrm{T07})]^2 M_{\star\mathrm{T07}}$. We compared the updated stellar masses with \citetalias{2007MNRAS.382..109T} stellar masses and found that the differences were small ($0.75 < \Mstar / M_{\star\mathrm{T07}} < 1.42$).
\item Individual galaxy spectra: taken from the DEEP2 DR4 survey. These spectra were obtained using three exposures of 20 min each with the multislit spectrograph DEIMOS \citep{2003SPIE.4841.1657F} on the Keck II telescope. The observations were performed under seeing conditions between 0.5 and 1.2 arcsec. Their spectral range is 6500--9300~\AA, and their spectral resolution $R$ is 5900 at 7800~\AA. They have a $\sim$2-\AA\ gap in the middle of the spectral range, owing to the multiCCD nature of the DEIMOS detector.
\end{enumerate}

\subsection{The T07-DEEP2 DR4 sample}

\begin{figure*}
  \includegraphics{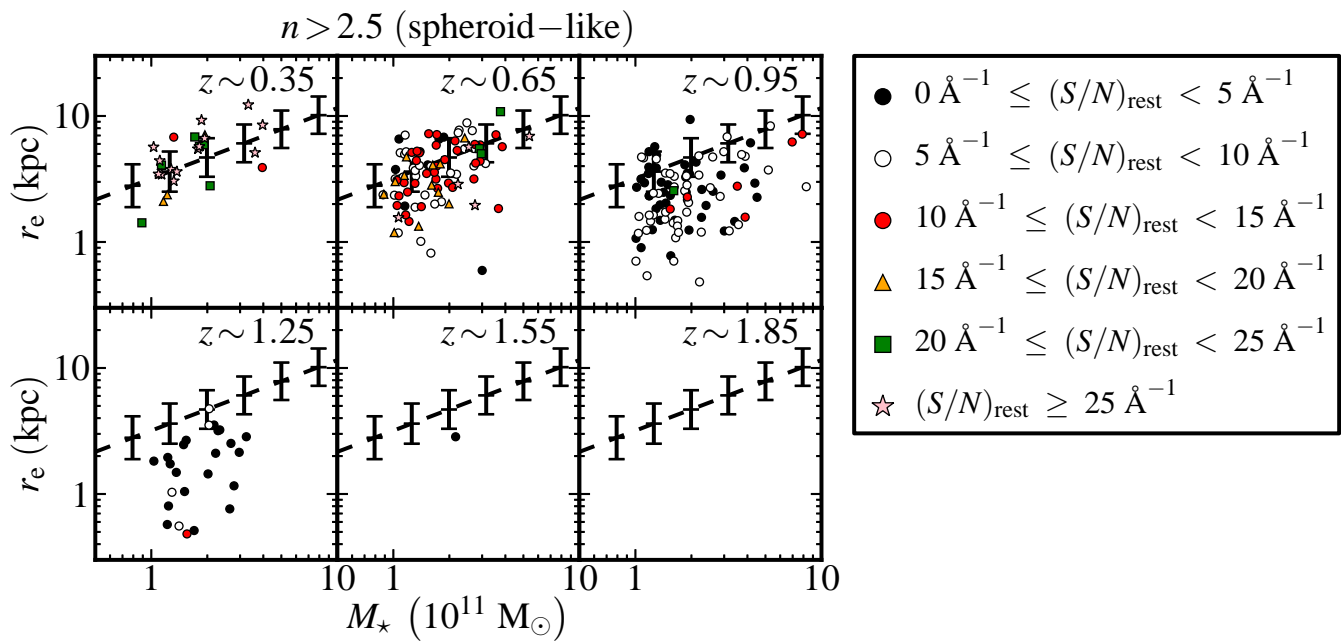}
  \caption{The stellar mass--size distribution of the spheroid-like galaxies ($n$ $>$ 2.5) with a good match between \citetalias{2007MNRAS.382..109T} and DEEP2 DR4 catalogues. The symbols represent the mean rest-frame signal-to-noise ratio of the DEEP2 DR4 spectra available for each object. The dashed line shows the relationship that the early-type galaxies follow in the nearby Universe \citep{2003MNRAS.343..978S}. Error bars over-plotted on the dashed line show the dispersion of this relationship. Stellar masses, radii and redshifts used in this figure were determined as detailed in Section~\ref{subsec:t07deep2-magnitudes}.}
  \label{fig:t07deep2-snr}
\end{figure*}

Fig.~\ref{fig:t07deep2-snr} shows the mean rest-frame signal-to-noise ratio of the spectra available for each target selected in Section~\ref{subsec:t07deep2-cross-matching} over their stellar mass--size distribution. Most galaxies are located in the three lower redshift bins. As we will use these spectra to make stacked spectra later on and to measure their velocity dispersions, we limit our present sample to targets selected in Section~\ref{subsec:t07deep2-cross-matching} that (i) have spectra with a signal-to-noise ratio greater than 5~\AA$^{-1}$, and (ii) have redshifts in the range $0.2 < z < 1.1$.

\subsection{Additional data: from SDSS up to high-redshift galaxies} \label{subsec:additional-data}

In Sections~\ref{sec:discrepancy} and \ref{sec:interpretation} we use additional data in the analysis of our results. We took some data of individual galaxies from the literature in order to explore whether our data are affected by potential biases arising from the stacking process employed in the \citetalias{2007MNRAS.382..109T}-DEEP2 DR4 sample. All these individual galaxies are massive ($\Mstar > 10^{11} \Msun$) and spheroid-like ($n > 2.5$). We used six nearby massive compact galaxies from \citet{2012MNRAS.423..632F}, four galaxies from \citet{2014ApJ...780..134S} at $z$ $\sim$ 0.5 \citep[these galaxies include the two galaxies from][]{2010ApJ...709L..58S}, and the four galaxies from \citet{2011ApJ...738L..22M} at $z$ $\sim$ 1. As detailed in the Introduction, a discrepancy between \Mstar\ and \Mdyn\ has been reported for all these galaxies. We have included dynamical and stellar masses from the compilation of masses and structural parameters for high-redshift individual galaxies carried out by \citet{2013ApJ...771...85V}. This compilation was made using data from \citet{2008ApJ...688...48V}, \citet{2006ApJ...644...30B}, \citet{2013ApJ...764L...8B}, \citet{2009ApJ...704L..34C}, \citet{2010ApJ...717L.103N}, \citet{2013ApJ...771...85V}, \citet{2012ApJ...755...26O}, \citet{2009Natur.460..717V} and \citet{2012ApJ...754....3T}, and it covers the redshift range from 0.9 to 2.2. We also took 32 massive spheroid-like galaxies from the recent work by \citet{2014ApJ...783..117B} with redshifts between 1.0 and 1.6.

We also use data from \citet{2009ApJ...696L..43C}, which include the results of the Sloan Digital Sky Survey NYU Value Added Galaxy Catalogue DR6 \citep{2005AJ....129.2562B,2007AJ....133..734B}, \citet{2005ApJ...631..145V,2008ApJ...688...48V}, \citet{2005A&A...442..125D} and \citet{2008A&A...482...21C}. These authors have provided us with the mean values of individual stellar masses and redshifts. Furthermore, \citet{2009ApJ...696L..43C} kindly provided us with tabulated values of mean sizes, redshifts and velocity dispersions used in their paper (except for their highest-redshift data point, where the mean velocity dispersion is the velocity dispersion on a stacked spectra). In this case, we assigned a mean dynamical mass to each data value, introducing the mean values of radius and velocity dispersion in equation~(\ref{eq:virial-cap}).

Finally, we added a sample of 53~571 galaxies from the NYU Value-Added Galaxy Catalog DR7 (hereinafter NYU SDSS sample, or the NYU sample of SDSS galaxies). These galaxies were selected using the following criteria: they are massive ($10^{11} \  \Msun < \Mstar < 10^{12} \  \Msun$), spheroid-like ($n$ $>$ 2.5), and close to the peak of the redshift distribution of the SDSS catalogue (0.05 $<$ $z$ $<$ 0.11). To avoid unreliable data, we applied extra restrictions to the SDSS sample: reliable velocity dispersions, physical radii and apparent sizes, namely $70 \  \kms < \sigma < 420 \  \kms$, 0.3 \kpc\ $<$ \re\ $<$ 30 \kpc, and \re\ $>$ 1 arcsec. The last condition is applied to ensure that the apparent sizes are larger than the average full width at half-maximum (FWHM) ground-based data. In addition, we applied the aperture correction for velocity dispersions given in equation (1) of \citet{2006MNRAS.366.1126C}, although our results are insensitive to this change.

\begin{figure*}
  \includegraphics{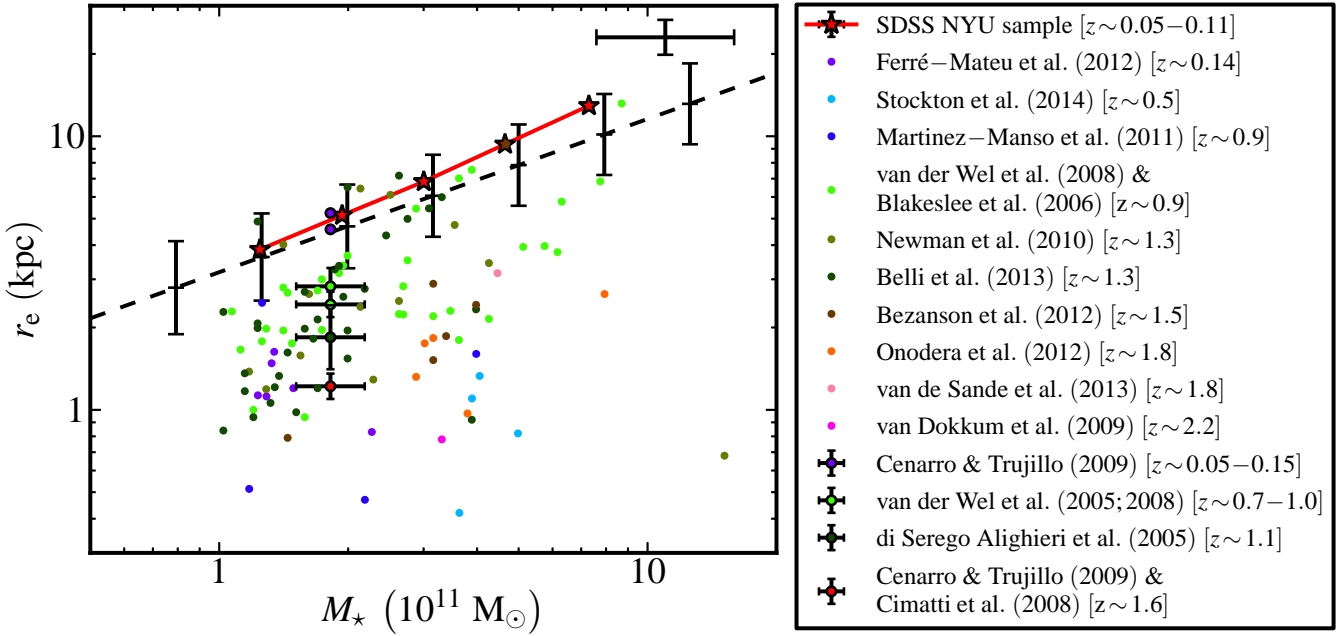}
  \caption{The stellar mass--size distribution of the additional data used in this paper described in Section~\ref{subsec:additional-data}. The symbols indicate the origin of each data set. Symbols with/without black edges represent the `average'/individual galaxies. For clarity, the NYU sample of SDSS galaxies has been plotted to show the mean properties of SDSS galaxies after a stellar-mass binning. The dashed line shows the relationship that the early-type galaxies follow in the nearby Universe \citep{2003MNRAS.343..978S}. Error bars over-plotted on the dashed line show the dispersion of this relationship. The error bar cross in the top right corner of the figure represents the mean error of the individual galaxies.}
  \label{fig:additional-r-m_star}
\end{figure*}

In Fig.~\ref{fig:additional-r-m_star} we show the distribution of these additional data in the stellar mass--size diagram.

\section{Processing of spectra and velocity dispersion measurements in the T07-DEEP2 DR4 sample} \label{sec:processing}

In order to obtain robust velocity dispersion measurements it is necessary to have a set of spectra with a high enough signal-to-noise ratio. This is achieved in our \citetalias{2007MNRAS.382..109T}-DEEP2 DR4 sample by using the spectral stacking technique. The steps followed by us in the processing of the spectra can be classified into three categories, which were applied in the following order: corrections to the individual spectra, stacking of spectra, and, finally, velocity dispersion measurements. The following subsections provide the details of each step.

\subsection{Corrections to the individual spectra}

The corrections applied to the individual spectra were as follows.

\begin{enumerate}
\item \emph{Throughput correction.} The spectra available in the DEEP2 DR4 data base are not relative-flux-calibrated. To obtain this calibration, we divided each spectrum by the throughput of the DEIMOS spectrograph \citep[see e.g.][]{2006ApJ...651L..93S}. We used the data published for an instrumental configuration with the gold 1200 l/mm grating and the OG550 filter.\footnote{The throughput data for several configurations have been made public at http://www.ucolick.org/$\sim$ripisc/results.html.} For the bluer regions of some spectra, the throughput data are not available, and hence we cut those parts of the spectra.
\item \emph{CCD-level relative calibration.} A visual inspection of the spectra shows that about 30 per cent of the spectra have a calibration problem between the two CCDs of the spectrograph. We corrected for this by applying a scaling factor to the red region of the spectrum. The scaling factors ranged from 0.6 to 1.5.
\item \emph{Pixel masking.} Some spectra have sharp variations at the CCD edges. Furthermore, two telluric absorption bands were identified (located at 6860--6922 and 7587--7714~\AA). These bands belong to the O$_2$ absorption spectrum \citep{1994MNRAS.267..904S}. These pixels were masked during the stacking process.
\item \emph{Flux normalization.} We carried out a flux normalization before the stacking process. For each redshift bin we used a different rest-frame wavelength region: 5500--5700~\AA\ for $0.2 < z < 0.5$; 4500--4700~\AA\ for $0.5 < z < 0.8$; and 4150--4250~\AA\ for $0.8 < z < 1.1$.
\end{enumerate}

\subsection{Stacking of spectra} \label{subsec:stacking-spectra}

We stacked our spectra in order to reach a signal-to-noise level that enabled us to measure the velocity dispersions with confidence. The stacking process was carried out as follows.

\begin{enumerate}
\item Redshift correction to each individual spectrum.
\item Interpolation of each spectrum to a common wavelength vector.
\item Averaging of spectra to be co-added. We used a non-weighted mean, but our essential results remain unchanged if we use a mean where weights are the mean signal-to-noise ratio per angstrom of each spectrum. We chose the non-weighted option because signal-to-noise weighting introduces biases towards the brightest galaxies.
\end{enumerate}

\begin{figure}
  \includegraphics{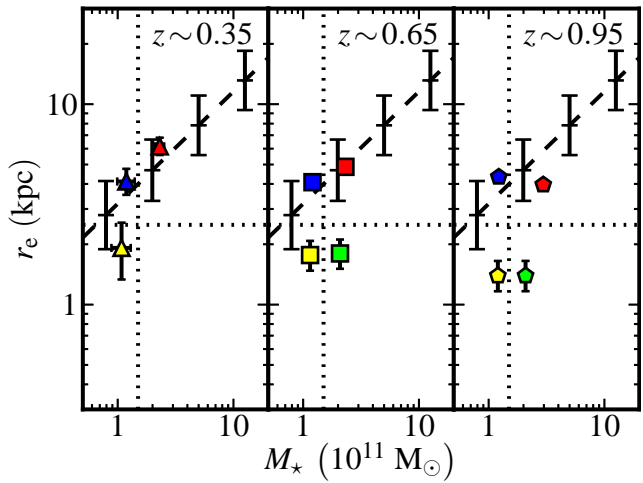}
  \caption{The stellar mass--size distribution of the stacked spectra described in Section~\ref{subsec:stacking-spectra}. Symbol shapes indicate different redshift bins, while colours indicate stellar masses and sizes. Dotted lines show the boundaries used to define each group of stacked spectra. The dashed line shows the relationship that the early-type galaxies follow in the nearby Universe \citep{2003MNRAS.343..978S}. Error bars over-plotted on the dashed line show the dispersion of this relationship.}
  \label{fig:stacked-r-m_star}
\end{figure}

We split our sample into three redshift bins ($0.2 < z < 0.5$, $0.5 < z < 0.8$ and $0.8 < z < 1.1$), and each redshift bin was further split into two stellar-mass bins and two effective radius bins. Boundaries were $1.5 \times 10^{11} \  \Msun$ and 2.5 kpc for mass and size, respectively. A stacked spectrum was built for each bin. Applying this procedure, we obtained 11 stacked spectra (the high-mass, small-size, low-redshift bin is empty). In Fig.~\ref{fig:stacked-r-m_star}, we show the mean stellar masses and radii corresponding to these stacked spectra.

\subsection{Velocity dispersion measurements} \label{subsec:sigma}

\begin{table*}
    \centering
    \begin{minipage}{175.5mm}
        \caption{Characteristics of the \citetalias{2007MNRAS.382..109T}-DEEP2 DR4 stacked spectra.}
        \label{tab:stacked-prop}
        \begin{tabular}{@{}lccd{1.3}cd{1.3}cd{2.0}d{2.1}cd{2.0}cd{2.0}@{}}
            \hline
            ID      & $\langle z \rangle$ & $\langle \Mstar \rangle$      & \multicolumn{1}{c}{$\Delta \langle \Mstar \rangle$} & $\langle \re \rangle$ & \multicolumn{1}{c}{$\Delta \langle \re \rangle$} & $\langle n \rangle$ & \multicolumn{1}{c}{$N_\mathrm{stack}$} & \multicolumn{1}{c}{$(S/N)_\mathrm{rest}$} & \sigmae & \multicolumn{1}{c}{$\Delta \sigmae$} & $\sigmae^\mathrm{MC}$ & \multicolumn{1}{c}{$\Delta \sigmae^\mathrm{MC}$}\\
                    &                     & ($10^{11} \  \Msun$) & \multicolumn{1}{c}{(dex)}                           & (kpc)                 & \multicolumn{1}{c}{(dex)}                        &                     &                                        & \multicolumn{1}{c}{(\AA$^{-1}$)}          & (\kms)  & \multicolumn{1}{c}{(\kms)}           & (\kms)                & \multicolumn{1}{c}{(\kms)}\\
            (1)     & (2)                 & (3)                           & \multicolumn{1}{c}{(4)}                             & (5)                   & \multicolumn{1}{c}{(6)}                          & (7)                 & \multicolumn{1}{c}{(8)}                & \multicolumn{1}{c}{(9)}                   & (10)    & \multicolumn{1}{c}{(11)}             & (12)                  & \multicolumn{1}{c}{(13)}\\
            \hline
            z0.35MR & 0.399               & 2.306                         & 0.04                                                         & 6.16                  & 0.007                                                     & 5.8                 & 13                                     & 33.9                                      & 205     & 3                                    & 204                            & 3\\
            z0.35mR & 0.351               & 1.188                         & 0.012                                                        & 4.11                  & 0.009                                                     & 6.0                 & 10                                     & 26.6                                      & 177     & 4                                    & 180                            & 3\\
            z0.35mr & 0.394               & 1.080                         & 0.03                                                         & 1.92                  & 0.014                                                     & 4.4                 & 3                                      & 14.9                                      & 203     & 7                                    & 201                            & 6\\
            z0.65MR & 0.688               & 2.316                         & 0.02                                                         & 4.87                  & 0.006                                                     & 5.5                 & 42                                     & 23.9                                      & 230     & 3                                    & 230                            & 3\\
            z0.65Mr & 0.706               & 2.080                         & 0.05                                                         & 1.80                  & 0.010                                                     & 4.7                 & 7                                      & 10.3                                      & 238     & 5                                    & 238                            & 6\\
            z0.65mR & 0.653               & 1.204                         & 0.009                                                        & 4.08                  & 0.011                                                     & 5.4                 & 21                                     & 12.3                                      & 190     & 5                                    & 186                            & 5\\
            z0.65mr & 0.688               & 1.147                         & 0.02                                                         & 1.77                  & 0.015                                                     & 5.3                 & 16                                     & 14.8                                      & 198     & 7                                    & 199                            & 6\\
            z0.95MR & 0.896               & 2.960                         & 0.06                                                         & 3.96                  & 0.015                                                     & 5.2                 & 21                                     & 11.1                                      & 230     & 8                                    & 237                            & 7\\
            z0.95Mr & 0.931               & 2.083                         & 0.03                                                         & 1.39                  & 0.02                                                      & 4.9                 & 17                                     & 10.4                                      & 231     & 9                                    & 228                            & 7\\
            z0.95mR & 0.817               & 1.223                         & 0.03                                                         & 4.33                  & 0.04                                                      & 4.2                 & 4                                      & 5.6                                       & 175     & 17                                   & 209                            & 20\\
            z0.95mr & 0.874               & 1.201                         & 0.014                                                        & 1.39                  & 0.03                                                      & 5.4                 & 12                                     & 8.0                                       & 198     & 11                                   & 207                            & 9\\
            \hline
        \end{tabular}\\
        Notes to Table~\ref{tab:stacked-prop}. (1) Identification associated with the stacked spectra. (2) Arithmetic mean redshift of the individual galaxies used in the stacked spectra. (3) Geometric mean stellar mass of the individual galaxies used in the stacked spectra. (4) Error of the geometric mean stellar mass. (5) Geometric mean effective (half-light) radius of the individual galaxies used in the stacked spectra. (6) Error of the geometric mean effective (half-light) radius. (7) Arithmetic mean S\'ersic index of the individual galaxies used in the stacked spectra. (8) Number of stacked individual galaxies. (9) Mean rest-frame signal-to-noise ratio per angstrom of the stacked spectra in the region where the velocity dispersion was measured. (10) Velocity dispersion measured in the stacked spectra. (11) Error of the velocity dispersion measured in the stacked spectra. (12) Mean velocity dispersion measured in 100 Monte Carlo realizations of the stacked spectra. (13) Standard deviation of the velocity dispersion measured in 100 Monte Carlo realizations of the stacked spectra.
    \end{minipage}
\end{table*}

In order to compute the velocity dispersion of our stacked spectra, we used the penalized pixel-fitting (\textsc{pPXF}) method of \citet{2004PASP..116..138C}. Details of the fitting procedure are as in \citet{2011MNRAS.417.1787F}. A relevant \textsc{pPXF} input parameter is the spectral library used to fit each spectrum. We used a set of 82 stars taken from the ELODIE library \citep{2007astro.ph..3658P} that cover a wide range of stellar parameters ($T_\mathrm{eff}$, [Fe/H], $\log(g)$), which allowed us to minimize the impact of template mismatch. Table~\ref{tab:stacked-prop} lists the characteristics of the \citetalias{2007MNRAS.382..109T}-DEEP2 DR4 stacked spectra, including the results from the velocity dispersion measurements.

In order to check the robustness of the error velocity dispersion errors, we performed a set of simulations, described in Appendix~\ref{ap:simulation}. We checked that the error estimations in this simulation were on average 1.8 times higher than the errors of the velocity dispersion measured in the stacked spectra in Table~\ref{tab:stacked-prop}, the ratios between these errors. As this simulation was developed based on an adverse single stellar population (of 10 Gyr), we conclude that the error estimations in Table~\ref{tab:stacked-prop} represent acceptable values. Furthermore, it is worth noting that even if the errors were underestimated by this factor the conclusions of the present paper would be unaltered.

In Appendix~\ref{ap:spectra}, we illustrate the \textsc{pPXF} fitting results for the stacked spectra to show the reliability of our spectral fits.

\section{Discrepancy between the dynamical and stellar masses} \label{sec:discrepancy}

We start by comparing the dynamical masses computed using equation~(\ref{eq:virial-cap}) with the stellar masses \Mstar.

\begin{figure}
  \includegraphics{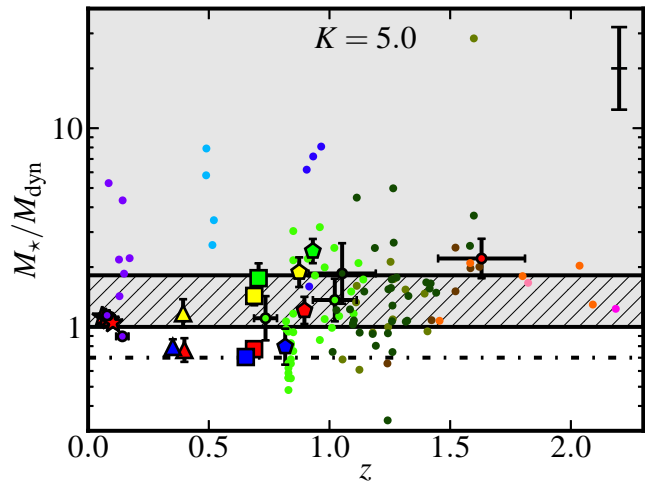}
  \caption{The $\Mstar / \Mdyn$ ratio as a function of redshift. The grey region represents the unphysical area where $\Mstar > \Mdyn$. The hatched region covers the sector where the discrepancy between the dynamical and stellar masses could be solved using a Chabrier IMF instead of a Salpeter IMF for the determination of the stellar masses. The dash--dotted line corresponds to a typical stellar/dynamical mass fraction \citep[0.7, see][]{2007ApJ...667..176G}. Symbols are as in Figs~\ref{fig:additional-r-m_star} and \ref{fig:stacked-r-m_star}. For clarity, the SDSS NYU sample has been plotted to show the mean properties of SDSS galaxies after a redshift binning. The error bar in the top right corner of the figure represents the mean $\Mstar / \Mdyn$ error of the individual galaxies. Redshift errors on individual galaxies are smaller than the symbols.}
  \label{fig:m_star/m_dyn-z}
\end{figure}

Most of the discrepancies between dynamical and stellar masses in the literature are reported in high-redshift studies. Thus, an immediate question arises: is this problem related to the redshift of the galaxies? Fig.~\ref{fig:m_star/m_dyn-z} shows that there is not a clear trend between $\Mstar / \Mdyn$ and redshift. This figure also shows that there are many galaxies in the forbidden region $\Mstar / \Mdyn > 1$ (grey area in the figure). It is clearly seen that our compact galaxies (yellow and green triangles, squares and pentagons) have larger $\Mstar / \Mdyn$ than our normal-sized galaxies (blue and red triangles, squares and pentagons). In addition, at low redshift the only conflicting values are the compact massive galaxies from \citet{2012MNRAS.423..632F}. Therefore, it is worthwhile to ask whether the \Mstar--\Mdyn\ problem is related to the compactness of the galaxies. This would explain why there is a trend to have more galaxies in the grey region at high redshifts in Fig.~\ref{fig:m_star/m_dyn-z}, as it could be connected with the strong size evolution of ETGs with redshift.

To check whether the source of the discrepancy between dynamical and stellar masses is the size of the objects, we need a criterion for measuring the compactness. In this work, we will employ the ratio $\re / \rShen$, where we define the function \rShen\ as the mean effective (half-light) radius of an ETG in the nearby Universe. Specifically, \rShen\ follows the equation
\begin{eqnarray}
\rShen \!\!\!\!\!\!\!\!\!\! & & \equiv 2.88 \times 10^{-6} \  \kpc \left( \frac{0.62 \  \Mstar}{\Msun} \right)^{0.56} \nonumber\\
       & & = 3.185 \  \kpc \left( \frac{\Mstar}{10^{11} \  \Msun} \right)^{0.56},
\end{eqnarray}
where we have considered the fitting result from \citet{2003MNRAS.343..978S}, and we have introduced the factor 0.62 to adapt the expression from these authors to a Salpeter IMF \citep{2009MNRAS.394..774L}.

\begin{figure}
  \includegraphics{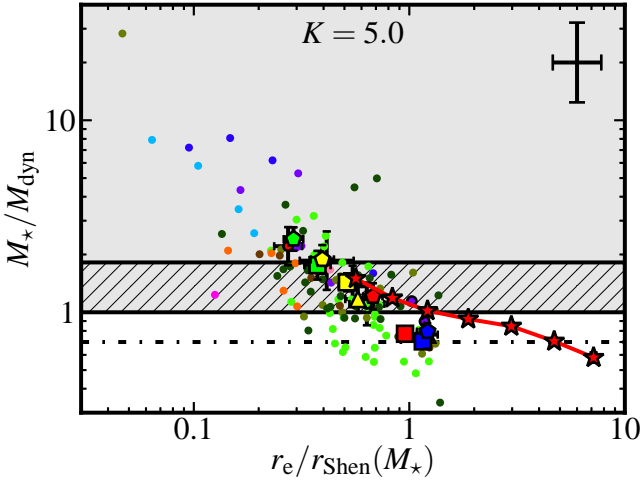}
  \caption{Correlation between the $\Mstar / \Mdyn$ ratio and the compactness indicator $\re / \rShen$. The symbols, regions, dash--dotted line and cross in the top right corner are as in Fig.~\ref{fig:m_star/m_dyn-z}. For clarity, the SDSS NYU sample has been plotted to show the mean properties of SDSS galaxies after a binning in $\re / \rShen$.}
  \label{fig:m_star/m_dyn-r/r_shen}
\end{figure}

Fig.~\ref{fig:m_star/m_dyn-r/r_shen} plots the fraction $\Mstar / \Mdyn$ versus the compactness indicator $\re / \rShen$. The correlation between the parameters represented in Fig.~\ref{fig:m_star/m_dyn-r/r_shen} is strong (the Spearman correlation coefficient is $-0.75$ for all data, and $-0.85$ for our data). To demonstrate the statistical significance of this correlation we performed two tests. First, we computed the two-tailed $p$-value for the Spearman correlation coefficient of all data and our data: $3 \times 10^{-24}$ and 0.001 respectively. The second test was to calculate the probability of the null result, namely whether the distribution of the data is compatible with no relation. To estimate that, we calculated the minimum $\chi^2$ value of a horizontal line fit to the data, and computed the probability of obtaining this or a higher $\chi^2$ in an uncorrelated distribution with the same degrees of freedom. We did this for the individual data and for \citetalias{2007MNRAS.382..109T}-DEEP2 DR4 data, obtaining values of $9 \times 10^{-13}$ and $3 \times 10^{-21}$ respectively. In addition, in this figure we can see that larger $\Mstar / \Mdyn$ values are found in the most compact galaxies (with several points around a factor of 8 and an extreme data point at $\sim$30).

The large ratio between the two masses cannot be justified in terms of the weakest aspect of the stellar mass determination: the IMF. Many authors have argued that there is a systematic variation in the IMF in ETGs \citep{2012Natur.484..485C,2012ApJ...760...71C,2013MNRAS.429L..15F}. However, their results would enable us to apply a correction factor to each galaxy of between 0.5 and 2. This is clearly insufficient to reconcile the data with a physically interpretable result. Consequently, one can ask whether the determination of the dynamical mass is the origin of the discrepancy. At this point, note that the determination of \Mdyn\ assumes the universality of the coefficient $K = 5.0$ of equation~(\ref{eq:virial-cap}). This assumption is based on the homology hypothesis. However, several authors have convincingly argued that ETGs are non-homologous systems \citep[see e.g.][]{2002A&A...386..149B,2006MNRAS.366.1126C}.

\begin{figure}
  \includegraphics{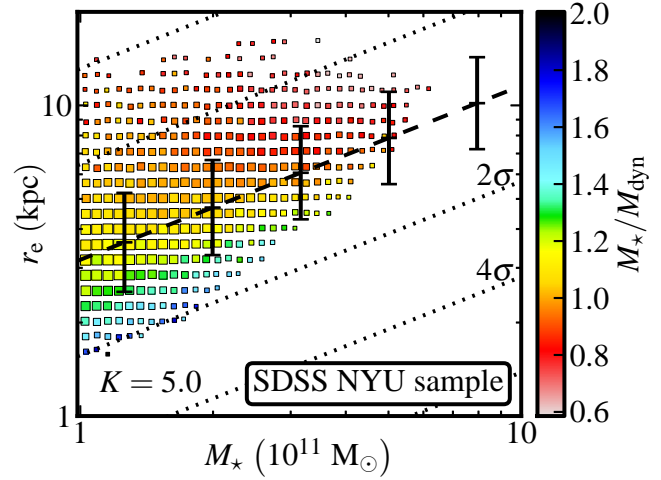}
  \caption{Dependence of the $\Mstar/\Mdyn$ ratio on the position in the \Mstar--\re\ diagram for galaxies in the NYU sample of SDSS galaxies. The data have been binned depending on stellar mass and radius. The colour of each symbol indicates the $\Mstar/\Mdyn$ ratio, while the size of each symbol scales with the number of galaxies in each bin. For clarity, bins with fewer than 10 galaxies have been omitted. The dashed line shows the relationship that the ETGs follow in the nearby Universe \citep{2003MNRAS.343..978S}. Error bars over-plotted on the dashed line show the dispersion of this relationship. Dotted lines are parallels to the dashed line spaced by twice the mean dispersion of the relationship from \citet{2003MNRAS.343..978S}.}
  \label{fig:m_star/m_dyn-in-m-for-nyu}
\end{figure}

We quantify non-homology effects in the next section. First, to reinforce our result that mass discrepancy correlates with compactness, we show in Fig.~\ref{fig:m_star/m_dyn-in-m-for-nyu}, using a large homogeneous sample of \textsl{nearby} galaxies, that the $\Mstar/\Mdyn$ ratio increases as the galaxies become increasingly compact. Fig.~\ref{fig:m_star/m_dyn-in-m-for-nyu} shows how the $\Mstar/\Mdyn$ ratio varies with the position on the \Mstar--\re\ diagram using the data from the NYU sample of SDSS galaxies. We note that contours that share the same $\Mstar/\Mdyn$ value tend to be parallel to the relationship that the ETGs follow in the nearby Universe \citep{2003MNRAS.343..978S} when $\re / \rShen < 1$. Above this relation, however, galaxies with similar $\Mstar/\Mdyn$ values seem to depend only on the $\re$ value and not on the mass.

\section{Interpretation of the discrepancy as a non-homology effect} \label{sec:interpretation}

In this section, we interpret the discrepancy between \Mstar\ and \Mdyn\ as a non-homology effect, and explain how the discrepancy can be solved using a variation of the $K$ coefficient in equation~(\ref{eq:virial-rsigma}).

\begin{figure}
  \includegraphics{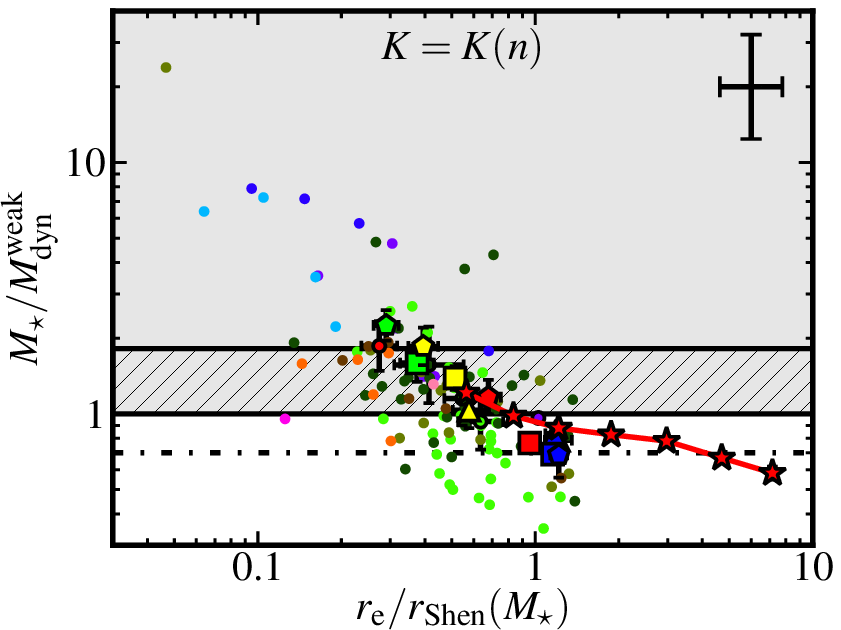}
  \caption{Correlation between the $\Mstar / \Mdyn^\mathrm{weak}$ ratio and the compactness indicator $\re / \rShen$. $\Mdyn^\mathrm{weak}$ refers to the dynamical mass estimated with equation~(\ref{eq:virial-rsigma}) and the weak homology hypothesis, namely a $K$ coefficient that depends on the S\'ersic index $n$. This figure uses the dependence of $K = K(n)$ measured by \citet{2006MNRAS.366.1126C}. Symbols, regions, the dash--dotted line and the cross in the top right corner are as in Fig.~\ref{fig:m_star/m_dyn-z}. For clarity, the SDSS NYU sample has been plotted to show the mean properties of SDSS galaxies after a binning in $\re / \rShen$.}
  \label{fig:m_star/m_dyn-weak-r/r_shen}
\end{figure}

The weak homology hypothesis proposes that deviations from homology are due to differences in the luminosity structure of galaxies. Indeed, different values of $K$ are expected from galaxies with different S\'ersic indices. \citet{2002A&A...386..149B} modelled this behaviour in their equation (11). \citet{2006MNRAS.366.1126C} quantified the dependence of $K$ on $n$ using their own sample (their equation 20):
\begin{equation} \label{eq:weak-homology}
K(n) = 8.87 - 0.831 n + 0.0241 n^2.
\end{equation}
The $K$ values from \citet{2006MNRAS.366.1126C} are higher than those of \citet{2002A&A...386..149B}, and hence come closer to resolving the mass discrepancy problem. We computed dynamical masses for our galaxies using equation~(\ref{eq:weak-homology}), and in Fig.~\ref{fig:m_star/m_dyn-weak-r/r_shen} plot $\Mstar / \Mdyn$ against the compactness indicator $\re / \rShen$. Fig.~\ref{fig:m_star/m_dyn-weak-r/r_shen} gives the mass discrepancy dependence assuming weak homology. Comparison with Fig.~\ref{fig:m_star/m_dyn-r/r_shen} shows that, while most points do shift down to lower values of $\Mstar / \Mdyn$ (median drop of $\log(\Mstar / \Mdyn) = 0.17$), the bulk of the galaxies remain in the unphysical region $\Mstar / \Mdyn > 1$. Hence, weak homology does not resolve mass discrepancy.

\begin{figure}
  \includegraphics{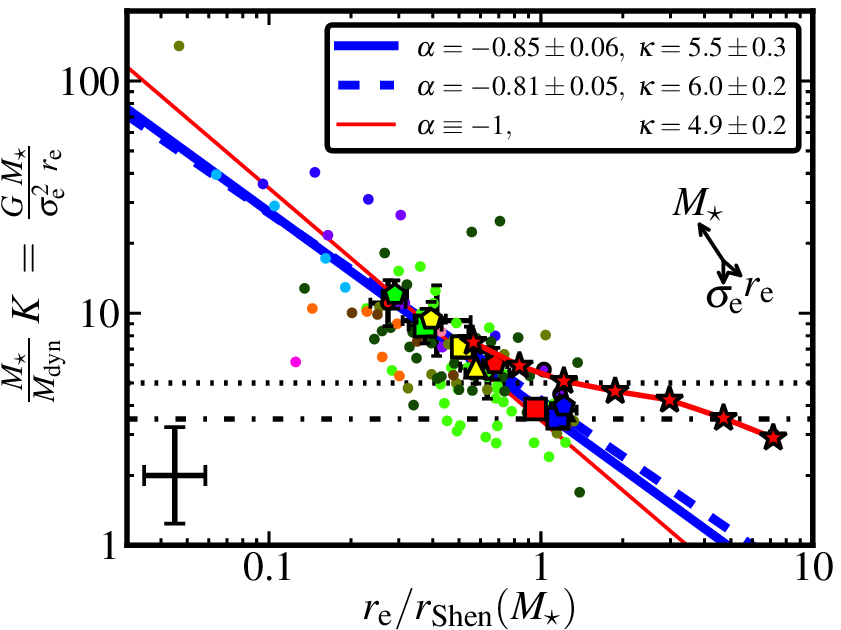}
  \caption{Dependence of $(\Mstar / \Mdyn) \  K$ on the compactness indicator $\re / \rShen$. The symbols are as in Figs~\ref{fig:additional-r-m_star} and \ref{fig:stacked-r-m_star}. The dash--dotted black line corresponds to a typical stellar/dynamical mass fraction \citep[0.7, see][]{2007ApJ...667..176G} and a $K$ value of 5.0 \citep{2006MNRAS.366.1126C}. The dotted black line corresponds to $(\Mstar / \Mdyn) \  K = 5.0$. Assuming $K = 5.0$, above this line we have the unphysical situation $\Mstar > \Mdyn$. The thick solid (dashed) blue line is a fit to all (our) data with the law $(\Mstar / \Mdyn) \  K = 0.7 \  \kappa \left( \re / \rShen \right)^\alpha$. The thin solid red line is a similar fit to all data on which the condition $\alpha \equiv -1$ has been imposed. For clarity, the SDSS NYU sample has been plotted to show the mean properties of SDSS galaxies after a binning in $\re / \rShen$. The error bar cross in the bottom left corner of the figure represents the mean error of the individual galaxies. The arrows in the right part of the figure indicate the shift that an individual point would have if we were to increase its \Mstar, \re\ or \sigmae\ by the mean error on these variables of the individual galaxies.}
  \label{fig:m_star/m_dyn*k-r/r_shen}
\end{figure}

Fig.~\ref{fig:m_star/m_dyn*k-r/r_shen} shows the dependence of $(\Mstar / \Mdyn) \  K$ on the compactness indicator $\re / \rShen$, where the first magnitude was calculated with the formula $(\Mstar / \Mdyn) \  K = (G \  \Mstar) / (\sigmae^2 \  \re)$. Assuming $K = 5.0$, we drew two horizontal lines in Fig.~\ref{fig:m_star/m_dyn*k-r/r_shen} for the cases $(\Mstar / \Mdyn) = 0.7$ \citep{2007ApJ...667..176G} and $(\Mstar / \Mdyn) = 1$ (i.e. no dark matter within the luminous body). Many objects lie above the $(\Mstar / \Mdyn) = 1$ boundary. This behaviour is not confined to high-redshift galaxies. Indeed, when $\re / \rShen < 1$, $\Mstar > \Mdyn$ also for galaxies in the SDSS NYU sample if we assume $K = 5.0$. The range of variation of $(\Mstar / \Mdyn) \  K$ in Fig.~\ref{fig:m_star/m_dyn*k-r/r_shen} is $\sim$1.5 dex, but an evolution from a stellar/dynamical mass fraction from 0.7 to 1 could only explain $\sim$0.15 dex; in other words, the observed range of $(\Mstar / \Mdyn) \  K$ is far from being fully explained by an increase of $(\Mstar / \Mdyn)$ for compact galaxies. The correlation of Fig.~\ref{fig:m_star/m_dyn*k-r/r_shen} has to be dominated by a variation of $K$ (its scatter may be given by galaxy-to-galaxy $\Mstar / \Mdyn$ variations owing to, for example, orientation effects or differences in rotational support). The only alternative to the conclusion that $K$ changes with compactness in Fig.~\ref{fig:m_star/m_dyn*k-r/r_shen} is that stellar mass determinations have large systematic errors \emph{that scale with compactness}. In this section we assume that this is not the case, although we will return to discuss this possibility in our conclusions. Therefore, Fig.~\ref{fig:m_star/m_dyn*k-r/r_shen} indicates a variation in the structure or dynamics of galaxies with compactness, which translates into a $K$ variation.

Can the so-called weak homology, where $K = K(n)$, explain the variation of $K$ shown in Fig.~\ref{fig:m_star/m_dyn*k-r/r_shen}? As shown in Fig.~\ref{fig:m_star/m_dyn-weak-r/r_shen}, this correction is much smaller than the value required to explain the range found in Fig.~\ref{fig:m_star/m_dyn*k-r/r_shen}. This means that non-homology effects owing to compactness are prevailing over weak homology effects.

\begin{figure}
  \includegraphics{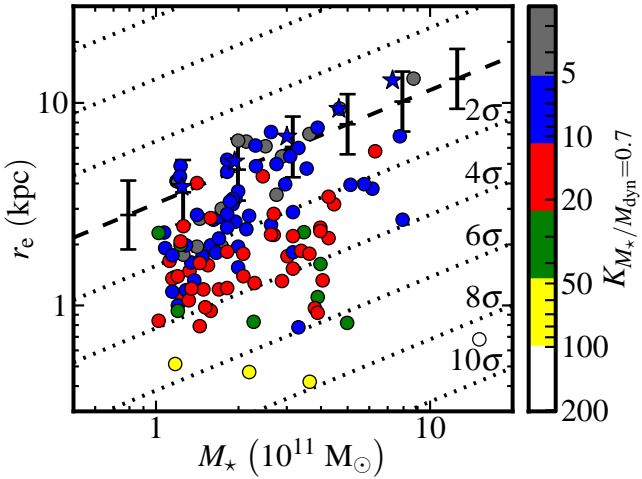}
  \caption{Variation of the $K$ coefficient in the stellar mass--size plane. A typical stellar/dynamical mass fraction in the computation of $K$ ($\Mstar / \Mdyn = 0.7$) has been assumed. The dashed line shows the relationship that the early-type galaxies follow in the nearby Universe \citep{2003MNRAS.343..978S}. Error bars over-plotted on the dashed line show the dispersion of this relationship. The dotted lines are parallels to the dashed line spaced by twice the mean dispersion of the relationship from \citet{2003MNRAS.343..978S}. For clarity, the SDSS NYU sample has been plotted to show the mean properties of SDSS galaxies after a stellar-mass binning and has been differentiated with star symbols.}
  \label{fig:k-in-m-r}
\end{figure}

\begin{figure*}
  \includegraphics{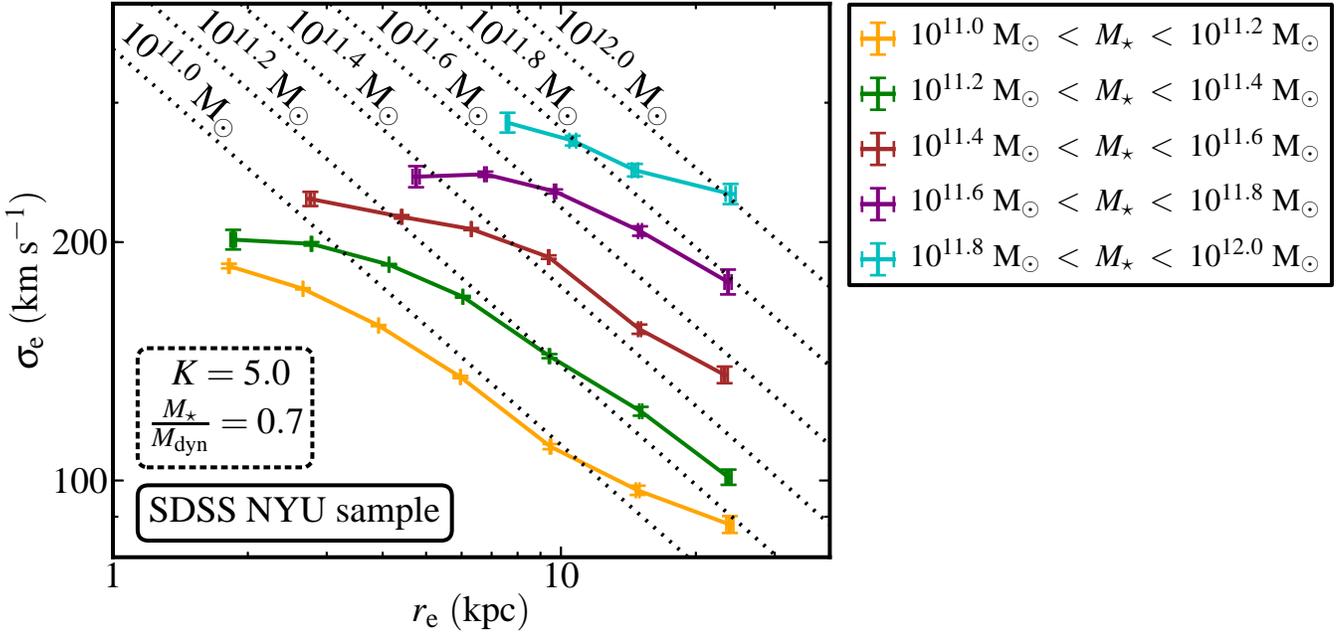}
  \caption{The relationship between \sigmae\ and \re\ the NYU sample of SDSS galaxies. The SDSS NYU data depart from the expectation of the virial theorem when the homology hypothesis is assumed (i.e. $\sigmae \propto \sqrt{\re}$). The dotted lines show the prediction for various stellar mass ranges. We have assumed $K = 5.0$ and $\Mstar / \Mdyn = 0.7$ for this prediction.}
  \label{fig:sigma-r}
\end{figure*}

To characterize the dependence of $K$ on galaxy compactness we fit a power law to the data in Fig.~\ref{fig:m_star/m_dyn*k-r/r_shen}; that is,
\begin{eqnarray} \label{eq:fit-k}
\frac{\Mstar}{\Mdyn} \  K \!\!\!\!\!\!\!\!\!\! & & = 0.7 \  \kappa \left( \frac{\re}{\rShen} \right)^\alpha \nonumber\\
                          & & = 0.7 \  \kappa \left( \frac{\re}{3.185 \  \kpc} \right)^\alpha \left( \frac{\Mstar}{10^{11} \  \Msun} \right)^{-0.56 \alpha},
\end{eqnarray}
where $\alpha$ and $\kappa$ are free parameters. $\alpha$ measures the logarithmic slope of $K$ with the compactness, while $\kappa$ is the value of $K$ for a galaxy that is located in the $z=0$ stellar mass--size relation with a typical stellar/dynamical mass fraction of $0.7$. Making an orthogonal distance regression, all data are fitted with $\alpha = -0.85 \pm 0.06$ and $\kappa = 5.5 \pm 0.3$. The use of only our own data gives $\alpha = -0.81 \pm 0.05$ and $\kappa = 6.0 \pm 0.2$. We have plotted these two fits with solid and dashed thick blue lines, respectively, in Fig.~\ref{fig:m_star/m_dyn*k-r/r_shen}. Given that the slope $\alpha$ in our fitting results is similar to $-1$, we also determined the $\kappa$ parameter fixing $\alpha \equiv -1$, obtaining $\kappa = 4.9 \pm 0.2$ (this last fit is plotted with a thin solid red line in Fig.~\ref{fig:m_star/m_dyn*k-r/r_shen}). It is remarkable that all $\kappa$ values are around the $K = 5.0$ value from \citet{2006MNRAS.366.1126C}. This result is, however, expected because the galaxies employed by those authors in their $K$ calibration were normal-sized ETGs in the $z=0$ stellar mass--size relation.

Once we have assumed that the variation in Fig.~\ref{fig:m_star/m_dyn*k-r/r_shen} results mainly from a variation in $K$, Fig.~\ref{fig:k-in-m-r} can be used to understand our results better. This figure shows the $K$ value for each galaxy assuming a fiducial stellar/dynamical mass fraction equal to 0.7 over the stellar mass--size plane. The geometrical meaning of our compactness indicator $\re / \rShen$ and its relationship with $K$ are clear in this figure: when the \Mstar--\Mdyn\ discrepancy is interpreted as a non-homology effect, it is the case that two galaxies are homologous (i.e. have the same $K$ value) if and only if their stellar masses and sizes are in the same parallel line to the relationship that ETGs follow in the nearby Universe. In addition, this figure shows the growth of $K$ as a power law of the departure from the $z=0$ stellar mass--size relation. In the figure the departure from the $z=0$ stellar mass--size relation is characterized using the dispersion of this relation (which has the value of 0.154 dex on \re\ axis). In addition, this figure emphasizes that the galaxies in the $z=0$ stellar mass--size relation are characterized by having the same value of the structural parameter $K$ ($K$ $\sim$ 5).

Another way of illustrating the departure from homology (i.e. the variation of $K$) as a function of the compactness is shown in Fig.~\ref{fig:sigma-r}. Using the NYU sample of SDSS galaxies, we plot the velocity dispersion of the galaxies versus their effective radii. We have split the sample into different stellar-mass bins. Fig.~\ref{fig:sigma-r} shows that galaxies with a similar stellar mass do not follow the homology hypothesis (i.e. their \sigmae\ is not proportional to $\sqrt{\re}$). Interestingly, the departure from homology is larger when the sizes of galaxies are smaller.

Finally, we provide a \emph{rough} approximation to the $K$ coefficient as a function of the compactness $\re / \rShen$ using the following expression:
\begin{eqnarray} \label{eq:k-prediction}
K \!\!\!\!\!\!\!\!\!\! & & \sim 6.0 \  \left( \frac{\re}{\rShen} \right)^{-0.81} \nonumber\\
  & & =    6.0 \  \left( \frac{\re}{3.185 \  \kpc} \right)^{-0.81} \left( \frac{\Mstar}{10^{11} \  \Msun} \right)^{0.45}.
\end{eqnarray}
This approximation is based on the fit to our data in Fig.~\ref{fig:m_star/m_dyn*k-r/r_shen}, and assumes a fiducial stellar/dynamical mass fraction of approximately 0.7 and a Salpeter IMF. We note that equation~(\ref{eq:k-prediction}) should be used only as an estimation within a factor of a few of the $K$ coefficient. If compact massive galaxies are strongly dominated by dark matter, namely if they have stellar/dynamical mass fractions lower than $\sim$0.3, equation~(\ref{eq:k-prediction}) has to be interpreted as a lower limit to the $K$ coefficient.

As a result of the dependence of $K$ on \re\ and \Mstar\  (equation~\ref{eq:k-prediction}) and using $\Mstar = 0.7 \  \Mdyn$, the dynamical mass adopts the following dependence on \re\ and \sigmae:
\begin{eqnarray} \label{eq:mdyn-prediction}
\Mdyn \!\!\!\! & \sim & \!\!\!\! \left[ 7.5 \  \left( \frac{\sigmae}{200 \  \kms} \right)^{1.6} \left( \frac{\re}{3 \  \kpc} \right)^{-0.65} \right] \frac{\sigmae^2 \re}{G} \nonumber\\
               & =    & \!\!\!\! \left( \frac{\sigmae}{200 \  \kms} \right)^{3.6} \left( \frac{\re}{3 \  \kpc} \right)^{0.35} 2.1 \times 10^{11} \  \Msun,
\end{eqnarray}
which highlights that the dependence on \Mdyn\ on \re\ and \sigmae\ strongly departs from the homology-virial theorem prediction (i.e. $\Mdyn \propto \sigmae^2 \re$, equation~\ref{eq:virial-rsigma} with $K$ independent of \sigmae\ and \re). We emphasize again that equation~(\ref{eq:mdyn-prediction}) should only be used as an estimate within a factor of a few of the dynamical mass. Regarding the two factors contributing to the mass discrepancy shown in Fig.~\ref{fig:m_star/m_dyn*k-r/r_shen}, equation~(\ref{eq:mdyn-prediction}) has corrected the dominant factor, namely compactness, by positing a shallow dependence of \Mdyn\ on radius ($\propto \re^{0.35}$). However, our formula does not address the other factor contributing to mass discrepancy, namely $\Mstar / \Mdyn$, because, by construction, equation~(\ref{eq:mdyn-prediction}) gives dynamical masses that fulfil $\Mstar / \Mdyn \sim 0.7$. Still, equation~(\ref{eq:mdyn-prediction}) should be useful because the expected departures from $\Mstar / \Mdyn = 0.7$ are small. Therefore, for a galaxy near to $\Mstar / \Mdyn \sim 0.7$, equation~(\ref{eq:mdyn-prediction}) approximates the dynamical mass. Conversely, when the baryon density is significantly lower than the dark matter density ($\Mstar / \Mdyn < 0.3$), equation~(\ref{eq:mdyn-prediction}) should be considered as a lower limit to the dynamical mass. It is possible to devise corrections to equation~(\ref{eq:mdyn-prediction}) if independent information on the behaviour of $\Mstar / \Mdyn$ is available.

A question that arises is whether in massive compact galaxies the ratio $\Mstar / \Mdyn$ (within $\sim$1 \re) significantly departs from the value $\Mstar / \Mdyn = 0.7$ found by \citet{2007ApJ...667..176G}. Two-dimensional spectroscopy of compact galaxies, followed by dynamical modelling, will provide the answer to this question. Our guess is that, if anything, $\Mstar / \Mdyn$ within the luminous effective radius is higher in an ultra-dense galaxy than in a normal galaxy, given the strongly dissipative processes that must occur in order for the galaxies to reach their high densities. Under this hypothesis, equation~(\ref{eq:mdyn-prediction}) should provide a good approximation to the dynamical mass.

\section{Discussion} \label{sec:discussion}

The family of ETGs may depart from homology (non-constant $K$ in equation~\ref{eq:virial-rsigma}) through differences in their internal structure, their internal dynamics and/or the contribution of dark matter to their gravitational potential. Structural non-homology is expected, given that the surface brightness profiles of ETGs show a range of S\'ersic indices $n$; \citet{2002A&A...386..149B} provide the dependence of $K$ on $n$. The internal dynamical structure of massive compact galaxies is also expected to differ from that of normal-sized massive galaxies: while the formation of massive compact galaxies must have involved a strongly dissipative collapse, normal-sized $z=0$ massive ellipticals owe their internal dynamics to a history of dissipationless mergers \citep[see e.g.][]{2011MNRAS.415.3903T}. Minor mergers, thought to be the dominant mechanism for the size evolution of massive compact galaxies, imply non-violent relaxation that leaves the remnant in a state far from equipartition per unit mass, whereas mixing is more pronounced in major mergers. A further clue that the internal dynamics of massive compact galaxies differs from that of normal-sized ellipticals is provided by the fact that these objects show elongated shapes \citep{2012MNRAS.423..632F,2014ApJ...780L..20T}; that is, they have disc-like shapes. At high redshifts, this has been confirmed by \citet{2012ApJS..203...24V} and \citet{2013MNRAS.428.1460B} among others. This fact may suggest an increase of the rotational contribution to the dynamical structure of these objects \citep{2013arXiv1305.0268B}.

We emphasize that our arguments leading to a conclusion of non-homology would be compromised if stellar mass estimates were shown to have large, systematic errors that scale with galaxy compactness. It remains, therefore, to seek clues on non-homology that are independent of stellar mass determinations, using deep two-dimensional spectroscopy of nearby massive compact galaxies to constrain Jeans or Schwarzschild modelling \citep[for example, it would be interesting to undertake a detailed study of the objects in][]{2012Natur.491..729V}.

\section{Conclusions} \label{sec:conclusions}

In what follows we summarize the main results of this paper.

First, the degree to which the ratio of stellar to dynamical masses is unphysical ($\Mstar / \Mdyn > 1$) is related to the compactness of the galaxies, not to redshift (Figs~\ref{fig:m_star/m_dyn-z} and \ref{fig:m_star/m_dyn-r/r_shen}). For most compact galaxies, the mass discrepancy is too large to be caused by the uncertainties in the IMF, arguably the weakest point in the determination of the total mass of a stellar population. Other uncertainties in the stellar mass determination cannot lead to a reconciliation of the two mass estimators either. The disagreement is too large to be explained by a variation in the dark matter fraction. Therefore, either (i) there exists an unknown large \emph{systematic} error in stellar masses linked to galaxy compactness, or (ii) there is a violation of the homology hypothesis in massive compact galaxies.

Second, when we interpret the \Mstar--\Mdyn\ discrepancy in terms of non-homology, namely as a variation of the coefficient $K$ from the virial theorem (equation~\ref{eq:virial-rsigma}), the value of $K$ reaches up to $\sim$40 for several compact galaxies (eight times greater than the value found by \citealt{2006MNRAS.366.1126C}). This strong variation in $K$ is well modelled as $K \propto (\re / \rShen)^{\alpha}$ with $\alpha \simeq -0.8$.

Finally, owing to the dependence of $K$ on \re\ and \Mstar, the dynamical mass scales with \sigmae\ and \re\ as $\Mdyn \propto \sigmae^{3.6} \re^{0.35}$, hence departing from homology and the virial theorem-based scaling $\Mdyn \propto \sigmae^2 \re$. Equation~(\ref{eq:mdyn-prediction}) provides an approximation to estimate \Mdyn\ for galaxies with $\Mstar / \Mdyn \sim 0.7$, and is a lower limit for galaxies strongly dominated by dark matter (i.e. $\Mstar / \Mdyn < 0.3$).

\section*{Acknowledgements}

We are grateful to the referee for his/her constructive comments, which helped to improve the manuscript. The authors thank A.~Ferr\'e-Mateu, A.~Vazdekis, J.~S\'anchez~Almeida, M.~Cappellari, E.~M\'armol-Queralt\'o, I.~G.~de~La~Rosa, A.~de~Lorenzo-C\'aceres, R.~C.~E.~van~den~Bosch, F.~Shankar, F.~Buitrago, A.~Stockton and the \emph{Traces of Galaxy Formation} group (http://www.iac.es/project/traces) for their comments during the development of this paper. LPdA is supported by the FPI Program by the Spanish Ministry of Science and Innovation. LPdA would like to thank I. Mart\'{\i}n-Navarro for fruitful discussions, and V. Al\'{\i}-Lagoa, L. Toribio San Cipriano and B. Gonz\'alez-Merino and C.~L\'opez-Sanjuan for comments that helped to improve the presentation of our results. JF-B acknowledges support from the Ram\'on y Cajal Program from the Spanish Ministry of Science and Innovation, as well as from the FP7 Marie Curie Actions of the European Commission, via the Initial Training Network DAGAL under REA grant agreement number 289313. This work was supported by the Programa Nacional de Astronom\'{\i}a y Astrof\'{\i}sica of the Spanish Ministry of Science and Innovation under the grants AYA2009-11137 and AYA2010-21322-C03-02. This research made use of \textsc{astropy}, a community-developed core \textsc{python} package for Astronomy \citep{2013A&A...558A..33A}. Funding for the DEEP2 Galaxy Redshift Survey was provided by NSF grants AST-95-09298, AST-0071048, AST-0507428 and AST-0507483 as well as by NASA LTSA grant NNG04GC89G. We acknowledge the usage in \textsc{pPXF} of the \textsc{mpfit} routine by \citet{2009ASPC..411..251M}.

 %***

\appendix

\section{Robustness of the error determination in the velocity dispersion measurements of the stacked spectra} \label{ap:simulation}

As an independent check of the accuracy of the velocity dispersion errors delivered by \textsc{pPXF} for our \citetalias{2007MNRAS.382..109T}-DEEP2 DR4 spectra, we ran a set of simulations with synthetic spectra to directly measure errors and their dependence on the signal-to-noise ratio and on velocity dispersion.

We first generated a grid of synthetic spectra as follows. We selected a spectrum of a single stellar population of age 10 Gyr and solar metallicity from the PEGASE library \citep {2004A&A...425..881L}. We chose this age because deriving velocity dispersions from spectra of this age is particularly challenging, and chose the solar metallicity as a good approximation to the metallicities of the galaxies under study. We degraded its spectral resolution to $R = 5900$, the same as the DEEP2 DR4 spectra, and degraded its velocity scale to 13 \kms\ pixel$^{-1}$, a mean value of the \citetalias{2007MNRAS.382..109T}-DEEP2 DR4 spectra at rest-frame. This adapted spectrum was convolved with Gaussian kernels to obtain 201 broadened versions of the spectrum corresponding to velocity dispersions from 140 to 260 \kms\ with a step of 0.6 \kms. With each one of these 201 spectra we built an additional 61 spectra, adding white noise to obtain spectra with a signal-to-noise ratio from 2.5 to 20.5 pixel$^{-1}$ (separated by 0.3 pixel$^{-1}$). We finally limited their spectral range from 4100 to 5800 \AA, a representative range of the \citetalias{2007MNRAS.382..109T}-DEEP2 DR4 spectra at rest-frame. We thus obtained a grid of 12261 spectra with 201 velocity dispersions and  61 signal-to-noise ratios.

\begin{figure}
  \includegraphics{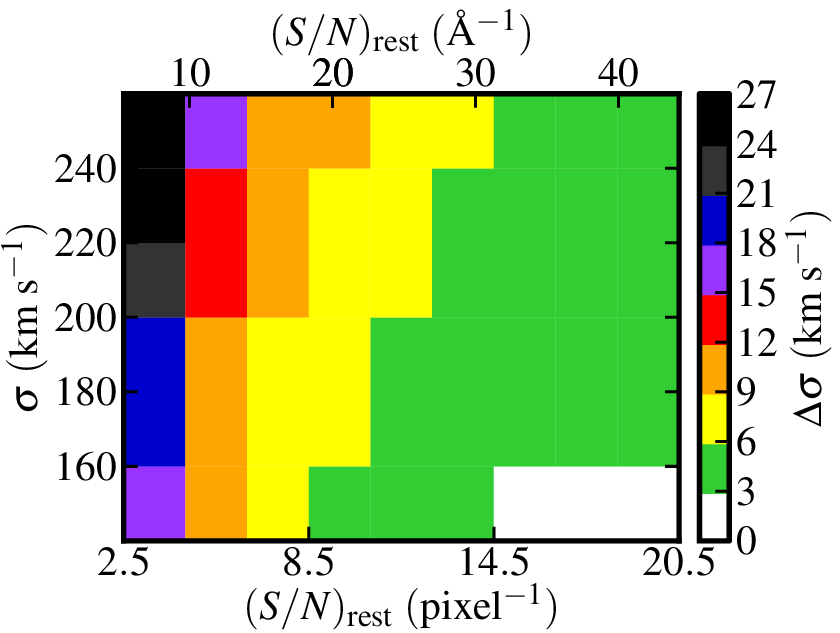}
  \caption{Grid of velocity dispersion deviations from the simulation described in Appendix \ref{ap:simulation} as a function of the signal-to-noise ratio and the input velocity dispersion. For clarity, we have binned the data according to their signal-to-noise ratio and $\sigma$ and computed the standard deviation.}
  \label{fig:simulation}
\end{figure}

\balance %***

In the next step, we measured the velocity dispersion at each point of the spectral grid; that is, we ran \textsc{pPXF} over 100 Monte Carlo realizations of each spectrum. In the \textsc{pPXF} executions, we used as input parameter the same 82-star spectral library that we used in Section~\ref{subsec:sigma} for fitting the \citetalias{2007MNRAS.382..109T}-DEEP2 DR4 stacked spectra. We computed the deviation at each point as the difference between the mean value of these measurements and the input velocity dispersion (known from the spectral grid generation step). We thus obtained a grid of velocity dispersion deviations at different velocity dispersions and signal-to-noise ratios. Fig.~\ref{fig:simulation} summarizes our results for this grid.

\begin{table}
    \centering
    \begin{minipage}{59mm}
        \caption{Comparison between different ways of estimating the velocity dispersion errors in the \citetalias{2007MNRAS.382..109T}-DEEP2 DR4 stacked spectra.}
        \label{tab:simulation}
        \begin{tabular}{@{}ld{2.0}d{2.0}d{2.0}}
            \hline
             ID     & \multicolumn{1}{c}{$\Delta \sigmae$} & \multicolumn{1}{c}{$\Delta \sigmae^\mathrm{MC}$} & \multicolumn{1}{c}{$\Delta \sigma^\mathrm{MC}_\mathrm{simul}$}\\
                    & \multicolumn{1}{c}{(\kms)}           & \multicolumn{1}{c}{(\kms)}                       & \multicolumn{1}{c}{(\kms)}\\
            (1)     & \multicolumn{1}{c}{(2)}              & \multicolumn{1}{c}{(3)}                          & \multicolumn{1}{c}{(4)}\\
            \hline
            z0.35MR & 3                                    & 3                                                & 4\\
            z0.35mR & 4                                    & 3                                                & 5\\
            z0.35mr & 7                                    & 6                                                & 9\\
            z0.65MR & 3                                    & 3                                                & 7\\
            z0.65Mr & 5                                    & 6                                                & 17\\
            z0.65mR & 5                                    & 5                                                & 11\\
            z0.65mr & 7                                    & 6                                                & 9\\
            z0.95MR & 8                                    & 7                                                & 16\\
            z0.95Mr & 9                                    & 7                                                & 17\\
            z0.95mR & 17                                   & 20                                               & 25\\
            z0.95mr & 11                                   & 9                                                & 19\\
            \hline
        \end{tabular}\\
        Notes to Table~\ref{tab:simulation}. (1) Identification associated with the stacked spectra (from Table~\ref{tab:stacked-prop}). (2) Error of the velocity dispersion measured in the stacked spectra (from Table~\ref{tab:stacked-prop}). (3) Standard deviation of the velocity dispersion measured in 100 Monte Carlo realizations of the stacked spectra (from Table~\ref{tab:stacked-prop}). (4) Error of the velocity dispersion obtained from the simulation described in Appendix~\ref{ap:simulation}.
    \end{minipage}
\end{table}

In the last step, we estimated the velocity dispersion errors of each \citetalias{2007MNRAS.382..109T}-DEEP2 DR4 spectrum. We computed these errors as the standard deviation of the grid velocity dispersion deviations inside a box of size 30 \kms\ $\times$ 2 pixel$^{-1}$ around the centre of \citetalias{2007MNRAS.382..109T}-DEEP2 DR4 spectral properties. In Table~\ref{tab:simulation}, we show the comparison between different ways of estimating the velocity dispersion errors in the \citetalias{2007MNRAS.382..109T}-DEEP2 DR4 stacked spectra, including the results obtained in this last step.

\onecolumn

\section{Fitting results for the stacked spectra} \label{ap:spectra}

\begin{figure*}
  \includegraphics[width=0.98\textwidth]{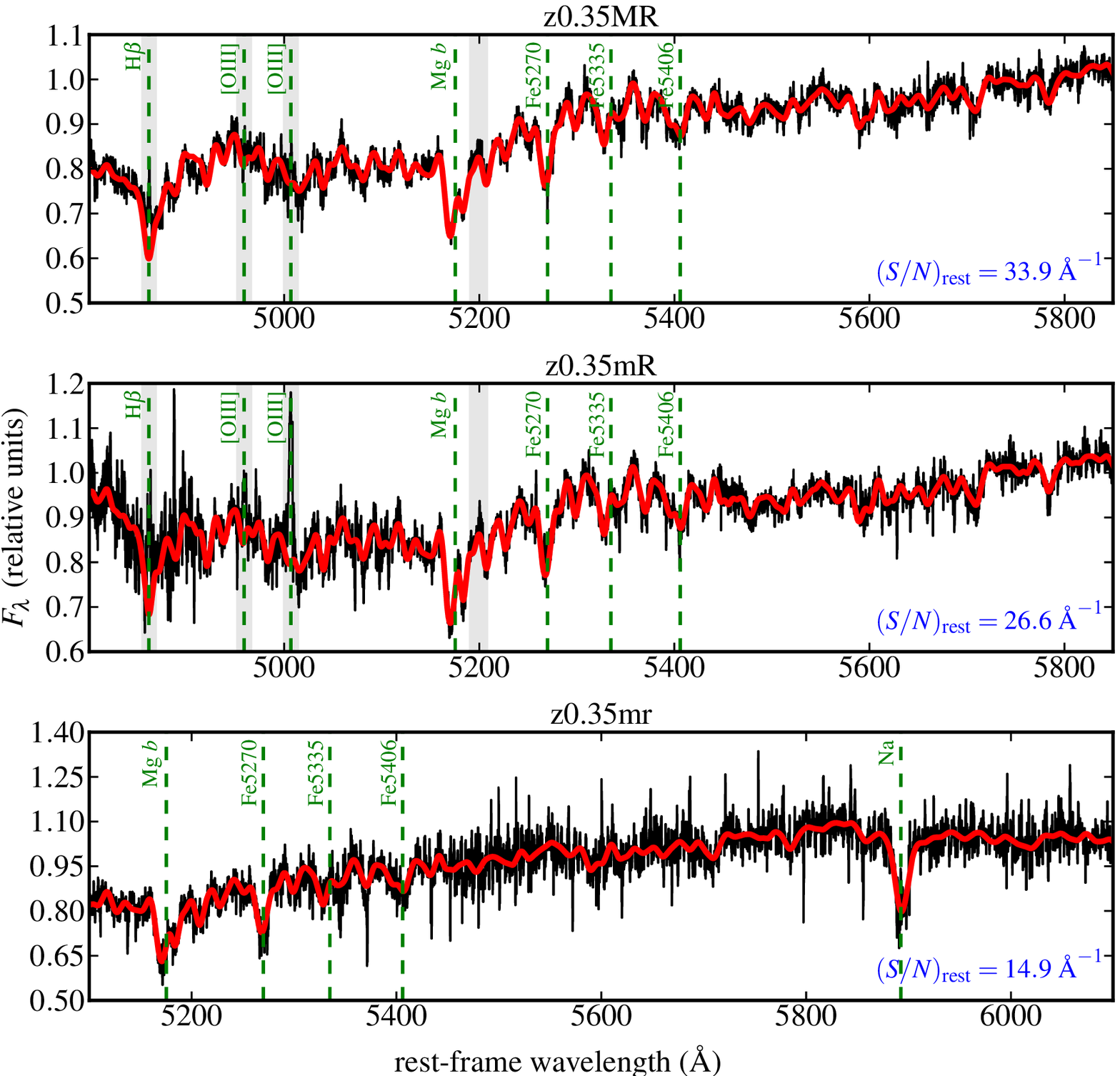}
  \caption{\textsc{pPXF} fitting results for the stacked spectra in the redshift bin $0.2 < z < 0.5$. The black solid lines represent our stacked spectra. Red thick solid lines are the \textsc{pPXF} fits. Masked regions in the \textsc{pPXF} fit are highlighted in grey. Some spectral features are marked with green dashed lines for visual clarity. Black labels above each panel are the stacked spectrum IDs from Table~\ref{tab:stacked-prop}. Blue labels in the bottom right corner of each panel indicate the mean rest-frame signal-to-noise ratio per angstrom of the stacked spectra in the region where the velocity dispersion was measured.}
  \label{fig:spectra-0.35}
\end{figure*}

\begin{figure*}
  \includegraphics[width=0.98\textwidth]{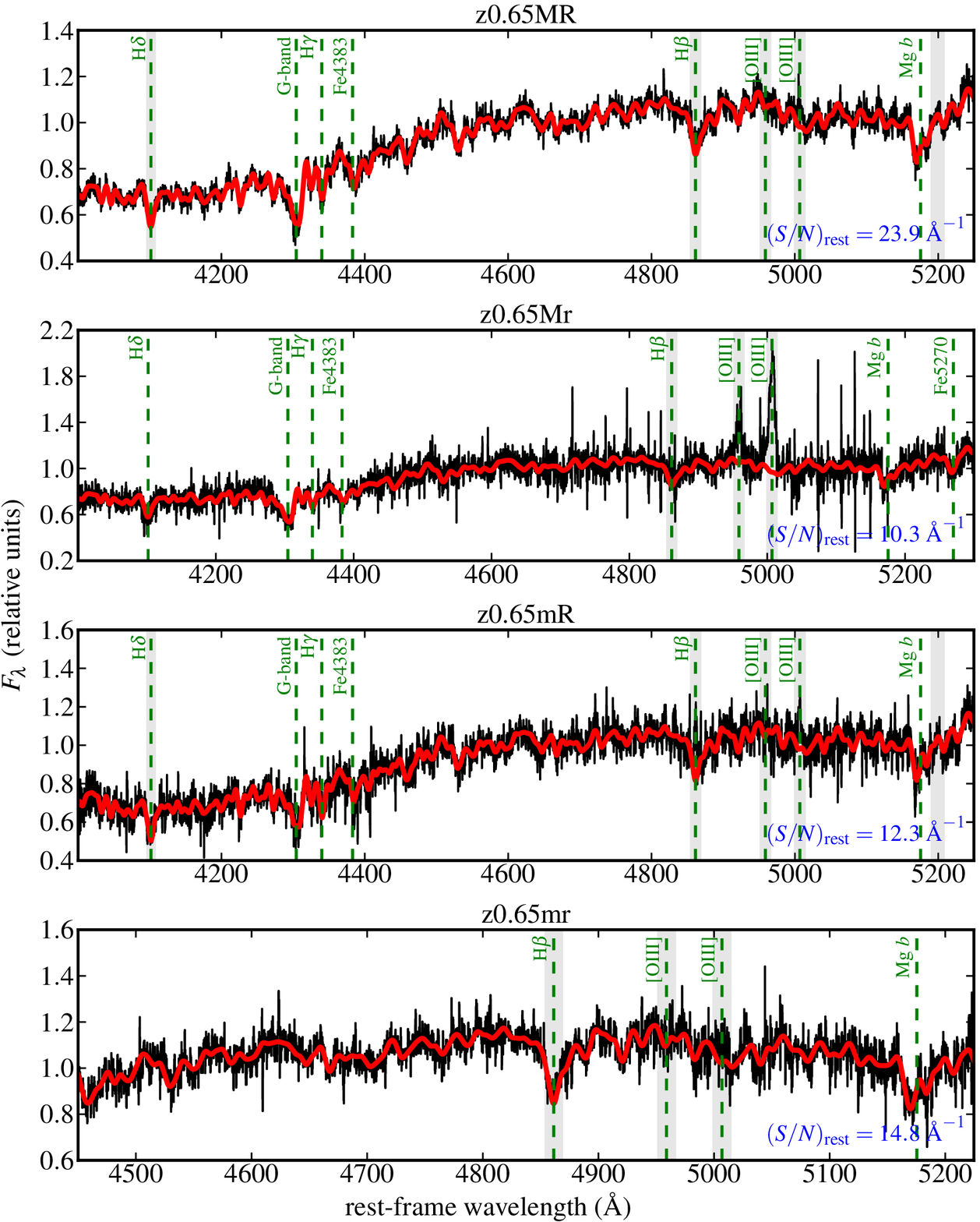}
  \caption{\textsc{pPXF} fitting results for the stacked spectra in the redshift bin $0.5 < z < 0.8$. The black solid lines represent our stacked spectra. Red thick solid lines are the \textsc{pPXF} fits. Masked regions in the \textsc{pPXF} fit are highlighted in grey. Some spectral features are marked with green dashed lines for visual clarity. Black labels above each panel are the stacked spectrum IDs from Table~\ref{tab:stacked-prop}. Blue labels in the bottom right corner of each panel indicate the mean rest-frame signal-to-noise ratio per angstrom of the stacked spectra in the region where the velocity dispersion was measured.}
  \label{fig:spectra-0.65}
\end{figure*}

\begin{figure*}
  \includegraphics[width=0.95\textwidth]{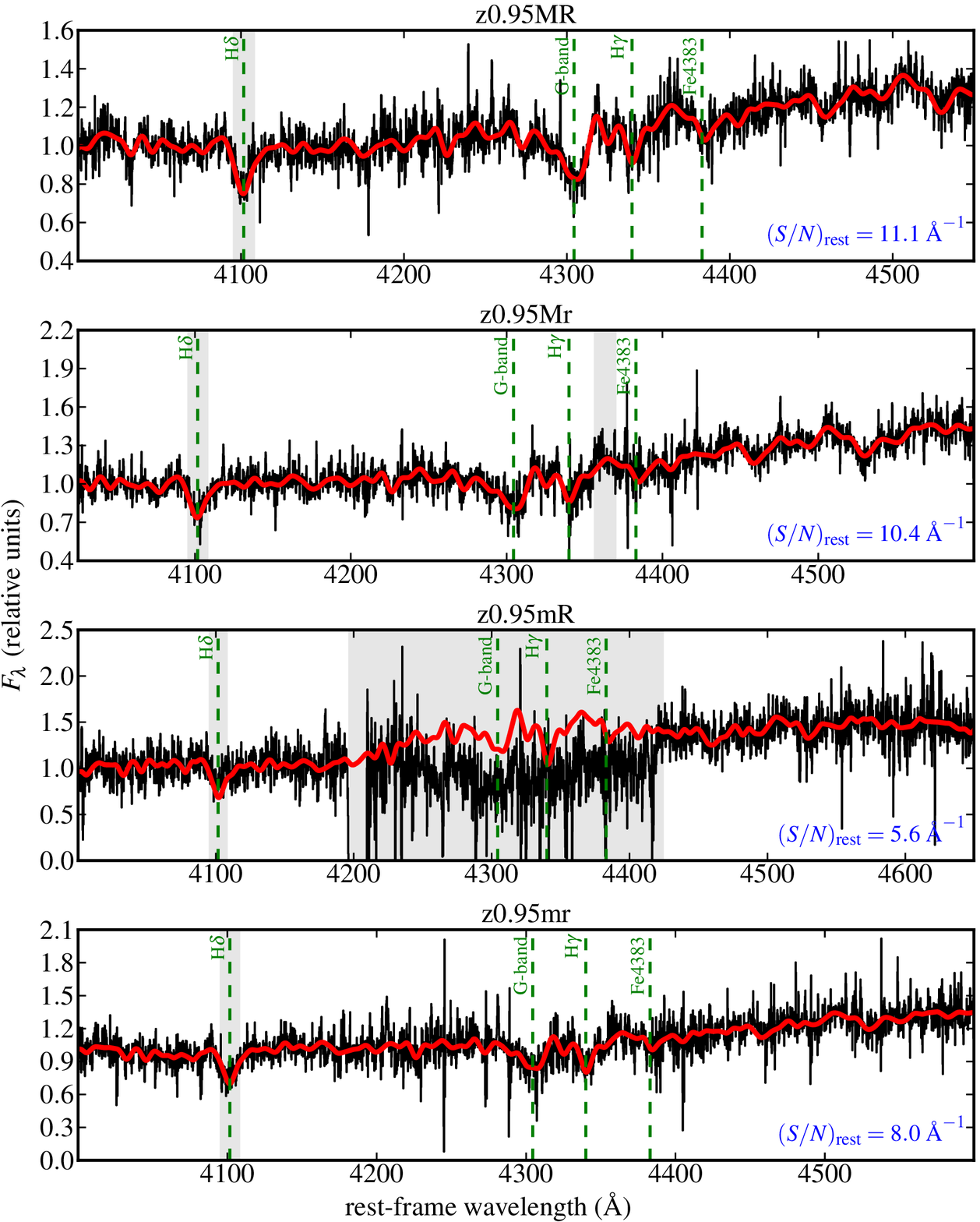}
  \caption{\textsc{pPXF} fitting results for the stacked spectra in the redshift bin $0.8 < z < 1.1$. The black solid lines represent our stacked spectra. Red thick solid lines are the \textsc{pPXF} fits. Masked regions in the \textsc{pPXF} fit are highlighted in grey. Some spectral features are marked with green dashed lines for visual clarity. Black labels above each panel are the stacked spectrum IDs from Table~\ref{tab:stacked-prop}. Blue labels in the bottom right corner of each panel indicate the mean rest-frame signal-to-noise ratio per angstrom of the stacked spectra in the region where the velocity dispersion was measured.}
  \label{fig:spectra-0.95}
  \label{lastpage} %***
\end{figure*}

We illustrate in Figs~\ref{fig:spectra-0.35}, \ref{fig:spectra-0.65} and \ref{fig:spectra-0.95} the \textsc{pPXF} fitting results for the stacked spectra described in Section~\ref{subsec:sigma} to show the accuracy of the \textsc{pPXF} spectral fitting.

\bsp

%\label{lastpage} %***

\end{document}